\documentstyle[12pt,epsf]{article}

\newcommand{\be}{\begin{equation}}
\newcommand{\ee}{\end{equation}}
\newcommand{\ben}{\begin{eqnarray}\displaystyle}
\newcommand{\een}{\end{eqnarray}}
\newcommand{\refb}[1]{(\ref{#1})}
\newcommand{\p}{\partial}
\newcommand{\sectiono}[1]{\section{#1}\setcounter{equation}{0}}

\newcommand{\II}{{\cal I}}
\newcommand{\CC}{{\cal C}}
\newcommand{\SS}{{\cal S}}
\newcommand{\VV}{{\cal V}}

\newcommand{\TT}{{\cal T}}

\newcommand{\NN}{{\cal N}}

\newcommand{\BB}{{\cal B}}

\newcommand{\QQ}{{\cal Q}}
\newcommand{\WW}{\Xi}   
\newcommand{\WWB}{{\langle \WW |}}

\newcommand{\wh}{\widehat}

\newcommand{\wb}{\bar}

\newcommand{\bmu}{{\bar\mu}}
\newcommand{\bnu}{{\bar\nu}}

\begin{document}
{}~
\hfill\vbox{\hbox{hep-th/0102112}\hbox{CTP-MIT-3082}
\hbox{PUPT-1976} \hbox{MRI-P-010201}
}\break

\vskip .8cm

\centerline{\large \bf Classical Solutions in String Field Theory}

\medskip

\centerline{\large \bf Around
the Tachyon Vacuum}

\vspace*{4.5ex}

\centerline{\large \rm Leonardo Rastelli$^a$, Ashoke Sen$^b$ and Barton
Zwiebach$^c$}

\vspace*{4.5ex}

\centerline{\large \it ~$^a$Department of Physics }

\centerline{\large \it Princeton University, Princeton, NJ 08540}

\centerline{E-mail:
        rastelli@feynman.princeton.edu}

\vspace*{2ex}

\centerline{\large \it ~$^b$Harish-Chandra Research
Institute\footnote{Formerly Mehta Research Institute of Mathematics
and Mathematical Physics}}

\centerline{\large \it  Chhatnag Road, Jhusi,
Allahabad 211019, INDIA}

\centerline{E-mail: asen@thwgs.cern.ch, sen@mri.ernet.in}

\vspace*{2ex}

\centerline{\large \it $^c$Center for Theoretical Physics}

\centerline{\large \it
Massachussetts Institute of Technology,}

\centerline{\large \it Cambridge,
MA 02139, USA}

\centerline{E-mail: zwiebach@mitlns.mit.edu}

\vspace*{5.0ex}

\centerline{\bf Abstract}
\bigskip

In a previous paper [hep-th/0012251] we proposed a simple class of
actions for string field theory around the tachyon vacuum. In this
paper we search for classical solutions describing D-branes of 
different dimensions using the ansatz that the solutions
factorize into the direct product of a matter state and a
universal ghost state. We find closed form expressions
for the matter state describing D-branes of all dimensions.
For the space filling D25-brane the state is 
the matter part of the zero angle wedge
state, the ``sliver'', built in [hep-th/0006240].  
For the other D-brane solutions the matter states  
are constructed using a solution generating
technique outlined in [hep-th/0008252]. The ratios of tensions of
various D-branes, requiring evaluation
of determinants of infinite dimensional matrices,
are calculated numerically and are in very good agreement
with the known results.

\vfill \eject

\baselineskip=16pt

\tableofcontents

\sectiono{Introduction and summary}  \label{s1} 

Cubic open string field theory \cite{WITTENBSFT} has turned out to be a
powerful tool in studying various conjectures~\cite{9902105,9904207}
about
tachyon condensation on bosonic
D-branes~\cite{KS,9912249,0002237,0002117,0003031,
0005036,0006240,0007153,
0008033,ellwood,0008053,0008101,0008252,0009105,0010190,
0011238,0101014,0101162}.
One aspect of the tachyon conjectures
that remains  to be confirmed is the expected
absence of physical open string excitations
around the tachyon vacuum.
In a previous
paper we proposed that a simple class of cubic
actions represent
string field theory built upon
the tachyon vacuum~\cite{0012251}.
As opposed to the conventional cubic SFT where the kinetic operator
is the BRST operator $Q_B$, here the kinetic operator $\QQ$ is
non-dynamical
and is built solely out of worldsheet ghost
fields.\footnote{A subset
of this class of actions was discussed previously in ref.\cite{HORO}.}
In this class of
actions
the absence of physical open string states around the vacuum is
manifest.
Gauge invariance holds in this class of actions, and therefore basic
consistency requirements are expected to be satisfied.

One major confirmation of the physical correctness of the proposed
actions would be the construction of
classical solutions which describe  the known D-brane
configurations.
An indirect argument for the existence of these solutions was given in
\cite{0012251} where it was also shown that under certain assumptions
the proposed action
reproduces in a rather nontrivial fashion
the  correct
ratios of tensions of D-branes of
different dimensions.

In this paper we give a
direct construction of
the classical solutions representing various D-branes and verify that
the ratios of their tensions 
agree with the known answer.
We use an ansatz where the solution $\Psi$ 
representing a D-brane has
a factorized form $\Psi_m\otimes\Psi_g$, with
$\Psi_m$ and $\Psi_g$ being
string fields built solely out of matter and
ghost operators respectively.\footnote{We wish to thank W.~Taylor
for emphasizing this factorization property to us.}
Such factorized form is clearly compatible with the structure
of the relevant string field equation since the kinetic operator
$\QQ$ does not mix matter and 
ghost sectors,\footnote{This is not true,  
of
course,
for the standard BRST operator.} and moreover, as is 
familiar, the star product also factors into 
the matter and ghost
sectors.
More explicitly, given string fields $A= A_m \otimes A_g$ and
$B= B_m \otimes B_g$, we have $A * B = (A_m *^m B_m ) \otimes (A_g *^g
B_g )$,
where $*^m$ and $*^g$ denote multiplication rules in
the
matter and ghost sectors respectively.
While the matter factor $\Psi_m$ is clearly different
for the various D-branes, we  assume that the ghost factor $\Psi_g$
is common to all the D-branes.

\smallskip
With this ansatz, and the specific form
of the action proposed in \cite{0012251}, the string field theory
equations of motion $\QQ \Psi + \Psi * \Psi =0$
factorizes into a matter part and a ghost part,
with the matter part yielding the equation $\Psi_m*^m\Psi_m =
\Psi_m$. We now ask how we can find solutions of this equation.
In fact, any solution of $\Psi * \Psi = \Psi$ where the ghost
number zero field $\Psi$ factors as $\Psi_m \otimes \Psi_g$
provides a solution of $\Psi_m*^m\Psi_m =
\Psi_m$.
There are at least two known translationally invariant solutions of
$\Psi * \Psi = \Psi$.
One solution is provided by the
identity string field $\II$, and the second is provided by the
``sliver", the zero angle wedge state $\WW$ 
constructed in
\cite{0006240}. This state was constructed in background
independent
language; it only requires the total Virasoro operators of the 
full
matter and ghost CFT. 
For reasons which will become clear later, 
we identify the matter part $\WW_m$ of
the sliver, 
with suitable normalization,
as the matter part of the solution describing the D25-brane.
More recently, Kostelecky and Potting
\cite{0008252} investigated
solutions of the equation $\Psi_m*^m\Psi_m = \Psi_m$ by using
an
explicit
representation of the $*$-product in terms of  Neumann
coefficients
for free (matter)
scalars.
In addition to the matter part of the
identity $\II$,
they found one nontrivial
solution
$\TT_m$.
Thus it is natural to ask: what
is the relationship between the matter component
$\WW_m$ of the sliver and the
state $\TT_m$ found in ref.\cite{0008252}?
We study this question using
the level
truncation scheme, and find very good evidence that the two
states are
really the same. 
Thus we can use either description for representing the D-25-brane. 
Our understanding of  the sliver 
$\WW_m$ shows that the solution describing
the D25-brane belongs to the universal subspace of the
state 
space \cite{9911116}, the space generated by the action of matter
Virasoro
generators and ghost oscillators on the SL(2,R) invariant
vacuum.
Furthermore, 
the sliver 
string field $\WW_m$
provides a ``simple closed-form" solution
for the matter factor representing the D25-brane in the SFT of
\cite{0012251}.
By ``closed form" we mean that the operational definition of
$\WW_m$ is explicit.
Even more, its geometrical meaning  is clear.
By ``simple" we mean that
the exact
calculation of $\WW_m$ 
to any given level requires only a
finite number
of
operations.\footnote{In
this vein one would say that the description
of this state as
$\TT_m$ \cite{0008252} is of closed form, as it is 
given by an explicit formula
in terms of an exactly calculable infinite 
dimensional matrix. 
The formula is not
simple in
that it involves inverses and square roots of this infinite matrix, so
even the finite level truncation of 
$\TT_m$ can only be
constructed approximately with finite number of operations.}

We then use the 
observations of \cite{0008252} to construct the
string field
$\Psi_m$ describing a lower dimensional
D-brane starting from the expression for the matter string field
representing the space filling D25 brane. The key point noted  
there is that the properties satisfied by the matter 
Neumann coefficients
$V^{rs}_{mn}$ ($m,n\geq 1$) 
that guarantee the existence of the
translational invariant solution are also satisfied by the extended
Neumann coefficients $V^{\prime rs}_{mn}$ ($m,n\geq 0$) 
defined by Gross and Jevicki 
\cite{gross-jevicki} by adding a new pair of oscillators to
represent the center of mass position and momentum operators.
Thus the same method 
used for generating translational invariant solution can be used to
generate lump solutions.
In implementing this procedure one requires 
the background dependent description $\TT_m$
of the D25 string field 
as a function of matter Neumann coefficients. 
We carry out the construction
thus obtaining ``closed-form" expressions
for the matter string fields representing lower dimensional branes.
The ratio of tensions of these
D-branes has an analytic expression in 
terms of the Neumann coefficients,
but explicit computation of this ratio involves evaluating determinants
of infinite dimensional matrices.
We calculate this
ratio in the level truncation using the known expression for the Neumann
coefficients.
While the convergence to
the answer is relatively slow as a function of the level $L$,
the relative simplicity of our expressions allows numerical
computations up to levels of the order of several thousands!
A fit of the data obtained at various levels suggests that corrections
vanish as inverse
powers of $\ln (L)$. The numerical results at large values of $L$ as
well
as an extrapolation of these results to $L=\infty$ via a fit using a
cubic
polynomial in $1/\ln(L)$ gives results very close to
the expected answer. We consider
these results
to be strong evidence for the correctness of the SFT we proposed in
\cite{0012251}.

Our concrete implementation of the procedure suggested in \cite{0008252}
actually finds families of solutions corresponding to lower
dimensional branes. 
The solutions  have gaussian profiles in the directions transverse
to the brane. We find that for a D-$(25-k)$ brane  
there is a $k$-parameter family of solutions, with the parameters
controlling the 
width of the lumps in different transverse directions. 
We believe that all these solutions are gauge equivalent.
This is necessary for the identification with D-branes,
since a physical D-brane has no moduli other than its position in the 
transverse space.
One indirect piece of evidence to this effect 
is that the ratios of tensions converge to the correct values for 
any solution in the family. Some more direct but still
incomplete arguments are given in appendix C.
In this context 
it will be interesting to explore if the width of the
lump in conventional string field theory, studied in ref.\cite{0005036},
changes when we use a gauge different from Siegel gauge.

Since besides the sliver   
state $\WW_m$, the identity
state $\II_m$ 
also squares to itself under the 
$*^m$ product, 
it is natural to ask why we identify the sliver  
and not the identity  as the solution representing the
D25-brane. While we do not have a concrete proof that $\II_m$
cannot be the matter part of the D25-brane solution, we offer the
following observations. First of all, 
as we have discussed, starting from
the sliver   state we can construct lump solutions of arbitrary
co-dimension
with correct ratios of tensions as expected of D-branes. If we  apply
the same procedure  to  
$\II_m$, we get back $\II_m$ and not a lower dimensional
brane. Thus, for example, 
 there is no obvious candidate for a 24-brane
solution with tension $2\pi$ times the tension associated with the state
$\II_m$, as would be expected of a D24-brane solution if
$\II_m$ represented the D25-brane. This clearly makes
$\WW_m$ a
much stronger candidate than $\II_m$ for the D25-brane
solution. $\II_m$ suffers from 
the  further complication that its normalization
properties are much worse  than those of 
$\WW_m$. Whereas the 
normalization of $\WW_m$ involves an infinite dimensional
determinant which is finite at least up to any given level (although it
could vanish as the level goes to infinity), 
the normalization of  
$\II_m$ involves  a determinant which vanishes at any finite level.

In the proposal of \cite{0012251} the explicit form of the kinetic
operator $\QQ$ was not fixed. In fact, we discussed two classes of
such operators. In the first class, exemplified by $\QQ = c_0$, the
operator
does not annihilate the identity string field $\II$. In the second
class, exemplified by $\QQ = c_0 + {1\over 2} (c_2 + c_{-2})$, the
operator
does  annihilate the identity string field $\II$. Both yield gauge
invariant
actions without physical open strings around the tachyon vacuum. A
proper
understanding of the ghost factor representing the D25 brane
(and in fact all other  D-branes,
since we assume that 
this factor is
universal)
would be expected to yield
some information on $\QQ$,  
since the ghost equation is of the form
$\QQ \Psi_g = - \Psi_g *^g \Psi_g$. Given that the matter part of the
string field for the D25-brane is the sliver 
state $\WW_m$, we expect
$\Psi_g$ to be closely related to the ghost part $\WW_g$ of the state.
Since $\WW_g$ is of ghost number zero and $\Psi_g$ must be of ghost
number one we conjecture that $\Psi_g = \CC \WW_g$ where $\CC$ is
a ghost number one operator built solely out of ghosts. It may turn out
that both $\QQ$ and $\CC$ are determined by
demanding the existence of a non-trivial solution to
the field equation.
Knowledge
of $\CC$ and $\QQ$ would amount to a complete specification of the SFT
action, and a complete knowledge of the string fields representing
D-branes.

\medskip
The rest of the paper is organized as follows. In section \ref{s2} we
discuss the factorization properties of the field equations, and give
the construction of the matter part of the  D25-brane solution in the
oscillator representation.
We also produce numerical evidence 
that this solution is identical to the matter part
of the sliver  state constructed
in ref.\cite{0006240}.
In section \ref{s3a} we construct the lump solutions, compute the
ratio of tensions of lump
solutions of different dimensions numerically and show that the result
is in very good agreement with the known results.
We conclude in section
\ref{s5} by listing some of the open questions. Appendix \ref{a1}
contains
a list of Neumann coefficients needed for our analysis.
Appendix \ref{a4} discusses the
transformation of the
3-string vertex when
we go from the momentum basis to the oscillator basis. It also contains
the precise  relationship between our variables and those
used in ref.\cite{gross-jevicki}, and some properties of the Neumann
coefficients which are important for our analysis. Appendix \ref{a2}
explores the possibility that a parameter appearing in the construction
of
the lump solution is a gauge artifact. In appendix \ref{a3} we derive
some properties of the sliver state.

\sectiono{Construction of the D25-brane solution} \label{s2} 

We begin with the string field theory action:
\be \label{e1}
\SS (\Psi) \equiv \,-\, {1\over g_0^2}\,\,\bigg[\, {1\over 2} \langle
\,\Psi \, ,
 \, \QQ\, \Psi
\rangle + {1\over 3}\langle \,\Psi \, , \, \Psi *
\Psi \rangle \bigg] \,,
\ee
where $\Psi $ is the string field represented by a state of ghost
number one
in the combined matter-ghost state space, $g_0$ is the open
string coupling constant, $\QQ$ is an operator made purely of ghost
fields and satisfying various requirements
discussed in
ref.\cite{0012251}, $\langle \, ,  \, \rangle$
denotes the BPZ inner product,
and $*$ denotes the usual $*$-product of the string
fields.
This action is supposed to describe the string field theory action
around
the tachyon vacuum. Although the action is formally background
independent, for practical computation ({\it e.g.} choosing a basis in
the
state space for expanding $\Psi $) we need to use a conformal  
field theory (CFT), and we take this to be the CFT describing the
D25-brane in flat space-time.

\subsection{Factorization property of the field equations} \label{ss21}

If \refb{e1} really describes the string field theory around the tachyon
vacuum, then the equations of motion of this field theory:
\be \label{e2}
\QQ \Psi = - \Psi * \Psi  \,,
\ee
must have a space-time independent solution describing the D25-brane,
and
also lump solutions of all codimensions describing lower dimensional
D-branes. We shall look for solutions of the
form:
\be \label{e3}
\Psi = \Psi_m \otimes \Psi_g\, ,
\ee
where $\Psi_g$
denotes a state obtained by acting with the ghost
oscillators on the SL(2,R) invariant vacuum of the ghost CFT, and
$\Psi_m$  is a
state obtained by acting with matter oscillators on the SL(2,R)
invariant
vacuum of the matter CFT.
Let us denote by
$*^g$ and $*^m$ the star product in the ghost and matter sector
respectively.
Eq.\refb{e2} then factorizes as
\be \label{e4}
\QQ \Psi_g = - \Psi_g *^g \Psi_g \,,
\ee
and
\be \label{e5}
\Psi_m = \Psi_m *^m \Psi_m\, .
\ee
Such a factorization is possible since $\QQ$ is made
purely of ghost operators. Note that we have used the freedom of
rescaling $\Psi_g$ and $\Psi_m$
 with $\lambda$ and $\lambda^{-1}$ to put
eqs.\refb{e4}, \refb{e5} in a convenient form.

In looking for the solutions describing D-branes of various dimensions
we shall assume that $\Psi_g$
remains the same for all solutions,
whereas $\Psi_m$
is different for different D-branes. Given two
static solutions of this kind, described by
$\Psi_m$ and $\Psi_m'$,
the ratio of the energy associated with these two
solutions is obtained by taking the ratio of the actions associated with
the two solutions. For a string field configuration satisfying the
equation of motion \refb{e2}, the action \refb{e1} is given by
\be \label{e6}
\SS |_\Psi = \,-\, {1\over 6 g_0^2}\,\, \langle
\,\Psi \, ,  \, \QQ\, \Psi\,
\rangle \, .
\ee
Thus with the ansatz \refb{e3} the action takes the form:
\be \label{e6a}
\SS |_\Psi = \,-\, {1\over 6 g_0^2}\,\, \langle
\,\Psi_g \, | \, \QQ\, \Psi_g \,\rangle_g \, \langle
\Psi_m|\Psi_m\rangle_m
\equiv K \langle \Psi_m|\Psi_m\rangle_m\, ,
\ee
where $\langle \, | \, \rangle_g$ and $\langle \, | \, \rangle_m$ denote
BPZ
inner products in ghost and matter sectors respectively.
$K= -(6g_0^2)^{-1} \langle \Psi_g|\QQ\Psi_g\rangle_g$
is a
constant factor calculated from the ghost sector which
remains the same for different solutions.
Thus we see that the ratio of the action associated with the two
solutions is
\be \label{e7}
{\SS |_{\Psi'}  \over \SS |_\Psi}  =
{\langle \Psi_m' | \Psi_m'\rangle_m \over \langle \Psi_m |
\Psi_m\rangle_m} \, .
\ee
The ghost part drops out of this calculation.

The analysis in the rest of this section will focus on the construction
of
a space-time independent solution to eq.\refb{e5} representing a
D25-brane. As pointed out in the introduction, there are two ways of
doing this. One method \cite{0006240} gives a description of this state
in
terms of matter Virasoro generators and the other method \cite{0008252}
describes this state in terms of the oscillators of the matter fields.
Since the second method can be generalized to describe lump solutions,
we
first describe this method in detail, and then compare this with the
first
description.

\subsection{A solution for the D25 brane}  \label{ss22}

Following ref.\cite{gross-jevicki,gj2,samuel} we
represent the
star product of two states $|A\rangle$ and $| B\rangle$ in the matter
CFT as\footnote{Whenever we use explicit operator representations
of the string product string fields will be denoted as kets or
bras as appropriate.}
\be \label{e8}
|A*^mB\rangle_3 = ~_1\langle A| ~_2\langle B| V_3\rangle  \,,
\ee
where the three string vertex $|V_3\rangle$ is given by
\be \label{e9}
|V_3 \rangle = \int d^{26}p_{(1)} d^{26}p_{(2)} d^{26}p_{(3)}
\delta^{(26)}(p_{(1)} + p_{(2)}
+ p_{(3)})
\exp (-  E)
|0, p\rangle_{123} \,,
\ee
and
\be
\label{e9p}
E =  {1\over 2} \sum_{{r,s}\atop m, n \ge 1}
 \eta_{\mu\nu} a^{(r)\mu\dagger}_m V^{rs}_{mn}
a_n^{(s)\nu\dagger} +
\sum_{{r,s}\atop n\ge 1}
 \eta_{\mu\nu} p_{(r)}^{\mu} V^{rs}_{0n}
a_n^{(s)\nu\dagger}
+{1\over 2}\sum_r\eta_{\mu\nu} p_{(r)}^{\mu}
V^{rr}_{00} p_{(r)}^{\nu}  \, .
\ee
Here $a^{(r)\mu}_m$, $a^{(r)\mu\dagger}_m$ are non-zero mode matter
oscillators
acting on the $r$-th string state
normalized so that
\be \label{e10}
[a^{(r)\mu}_m, a^{(s)\nu\dagger}_n] =
\eta^{\mu\nu}\delta_{mn} \delta^{rs},\qquad m,
n\ge 1\, .
\ee
$p_{(r)}$ is the 26-component momentum of the $r$-th string, and $|0,
p\rangle_{123}\equiv |p_{(1)}\rangle\otimes |p_{(2)}\rangle\otimes
|p_{(3)}\rangle$ is the
tensor product of Fock vacuum of the three strings, annihilated by the
non-zero mode annihilation operators $a^{(r)\mu}_m$, and eigenstate of
the
momentum operator of the $r$th string with eigenvalue $p_{(r)}^{\mu}$.
$|p\rangle$ is normalized as
\be \label{enorm}
\langle p|p'\rangle = \delta^{26}(p+p')\, .
\ee
The
coefficients $V^{rs}_{mn}$ for $0\le m, n<\infty$ can be calculated by
standard methods~\cite{gross-jevicki, samuel} and have been
given in appendix \ref{a1}.\footnote{In our conventions we take
$\alpha'=1$.} Some properties of $V^{rs}_{mn}$ have been discussed in
appendix \ref{a4}. Since we are interested in this section in
space-time translational invariant
solutions, we can ignore the
momentum
dependent factors in the vertex.
and the relevant form of $E$ is:
\be \label{vsimp}
E =   {1\over 2} \sum_{{r,s}}
 \eta_{\mu\nu} a^{(r)\mu\dagger}\cdot V^{rs}
\cdot a^{(s)\nu\dagger}
\,,
\ee
where the dots represent sums over mode numbers,
and $V^{rs}_{mn}$ for $m,n\geq 1$ is written as the
$V^{rs}$ matrix.
For the analysis of lumps, however, 
 we will need the full vertex.

Some appreciation of the  properties reviewed in
Appendix \ref{a4} is necessary  for
structural reasons.  Equation \refb{e11}, in particular, gives
\be \label{e11p}
V^{r\, s} = {1\over 3} ( C + \omega^{s-r} U + \omega^{r-s} \wb U)
\, ,
\ee
where $\omega=e^{2\pi i/3}$,  
$U$ and $C$ are regarded as matrices with
indices
running
over $m,n \geq 1$,
\be
\label{cdef}
C_{mn} = (-1)^m \delta_{mn}, \quad m,n\geq 1 \, ,
\ee
and $U$ satisfies (\refb{e133})
\be \label{e133p}
\wb U \equiv U^*= C U C, \qquad U^2 = \wb U^2 = 1, \qquad U^\dagger =
U\,
,\quad \wb U^\dagger = \wb U\, .
\ee
The superscripts $r,s$ are defined mod(3), and \refb{e11p} manifestly
implements the cyclicity property $V^{rs}=V^{(r+1)(s+1)}$. 
Also note the transposition 
property  
$(V^{rs})^T = V^{sr}$. 
Finally,
eqs.\refb{e11p}, \refb{e133p} allow one to show that
\be \label{e21aa1}
[C V^{rs}, C V^{r' s'}] = 0 \quad \forall \quad r, s, r', s',
\ee
and
\ben \label{e21aa2}
&&(CV^{12})(CV^{21}) = (CV^{21})(CV^{12}) = (CV^{11})^2 - CV^{11}\, ,
\nonumber \\
&&(CV^{12})^3 + (CV^{21})^3 = 2 (CV^{11})^3 - 3 (CV^{11})^2 + 1\, .
\een
Equations \refb{e11p} up to \refb{e21aa2}
are all that we shall need 
to know about the matter part of the relevant star product
(as given in eqs. \refb{e8}, \refb{e9} and \refb{vsimp}) to construct
the
translationally invariant solution. In fact, since \refb{e21aa1},
\refb{e21aa2}
follow from \refb{e11p} and \refb{e133p},
these two equations are really all that is strictly needed.
Such structure will reappear in the next section with matrices
that also include $m=0$ and $n=0$ entries, and thus will
guarantee the existence
of a solution constructed in the same fashion as the solution to be
obtained below.

\bigskip
We are looking for a space-time independent solution
of eq.\refb{e5}.
The strategy of
ref.\cite{0008252} is to take a trial solution of the
form:\footnote{We caution
the reader that although
in this section and in section \ref{s3a}
we shall follow the
general strategy
described in \cite{0008252}, our explicit formul\ae\
differ from theirs in several instances.}
\be \label{e15}
|\Psi_m\rangle = \NN^{26} \exp\Big(-{1\over 2}
\,\eta_{\mu\nu}\sum_{m,n\ge 1}S_{mn}\, a^{\mu\dagger}_m
a^{\nu\dagger}_n\Big) |0\rangle \,,
\ee
where $|0\rangle$ is the SL(2,R) invariant vacuum of the matter CFT,
$\NN$ is a normalization factor, and $S_{mn}$ is an infinite dimensional
matrix with indices $m,n$ running from 1 to $\infty$. We shall take $S$
to be twist invariant:\footnote{Due to this property
the BPZ conjugate of the state $|\Psi_m\rangle$ is the same as its
hermitian conjugate. Otherwise we need to keep track of extra $-$ signs
coming from the fact that the BPZ conjugate of $a_m^\dagger$ is
$(-1)^{m+1} a_m$.}
\be
\label{tinv}
CSC=S \, .
\ee
We shall check in the end that the solution constructed
below is indeed twist invariant.

If we define
\be \label{e16}
\Sigma = \pmatrix{S & 0\cr 0 & S},
\qquad \VV = \pmatrix{V^{11}
& V^{12}
\cr V^{21} & V^{22}}\, ,
\ee
and 
\be \label{e18}
\chi^{\mu \, T} = \pmatrix{ a^{(3)\mu\dagger} V^{31} \,, &
a^{(3)\mu\dagger}V^{32}}, \qquad
\chi^{\mu} = \pmatrix{ V^{13}
a^{(3)\mu\dagger} \cr
V^{23} a^{(3)\mu\dagger}}
\, ,
\ee
then using eqs. \refb{e8}, \refb{e9},
\refb{vsimp} we get 
\ben \label{e17}
|\Psi_m * \Psi_m\rangle_3 &=& \NN^{52} \,{\rm det} \{ (1 - \Sigma
\VV)^{-1/2}\}^{26} \nonumber \\
&\times&\hskip-8pt
\exp \Bigl[ -{1\over 2}\,\eta_{\mu\nu} \{
\chi^{\mu \, T} [(1 - \Sigma \VV)^{-1} \Sigma]
\chi^{\nu} + a^{(3)\mu\dagger} \cdot V^{33} \cdot
a^{(3)\nu\dagger}\} \Bigr]|0\rangle_3\, . 
\een
In deriving eq.\refb{e17} we have used
the general
formula \cite{0008252}
\ben \label{egen}
&& \langle 0 | \exp\Big(\lambda_i a_i -{1\over 2} P_{ij} a_i a_j\Big)
\exp\Big(\mu_i a^\dagger_i -{1\over 2} Q_{ij} a^\dagger_i
a^\dagger_j\Big)
|0\rangle
\nonumber \\
=&&\hskip-13pt \det(K)^{-1/2} \exp\Big(\mu^T\, K^{-1} \lambda -{1\over
2}
\lambda^T \,Q \, K^{-1} \lambda - {1\over 2}\mu^T\, K^{-1}P\mu\Big)\,
,\,\,
K\equiv 1- PQ\,.\,
\een
In using this formula we took the $a_i$ to be the list of oscillators
$(a^{(1)}_m, a^{(2)}_m)$ with $m\geq 1$. 
\refb{e17} then follows  
from \refb{egen} by identifying $P$ with $\Sigma$,
$Q$ with $\VV$, $\mu$ with $\chi$ and setting $\lambda$ to 0.

Demanding that
the exponents in
the
expressions for
$|\Psi_m\rangle$ and $|\Psi_m*\Psi_m\rangle$, 
given in eqs.\refb{e15} and
\refb{e17} respectively, match, we get
\be \label{expon}
S = V^{11} + (V^{12}\,,  V^{21}) (1 - \Sigma\VV)^{-1} \Sigma
\pmatrix{V^{21} \cr V^{12}}\, ,
\ee
where we have used the cyclicity property of the $V$ matrices and
the mod 3 periodicity of the indices $r$ and $s$ to write the equation
in a convenient form.
To proceed, we assume that
\be \label{econstraint}
[C S, C V^{rs}] = 0 \quad \forall \quad r, s\, .
\ee
We shall check later that the solution obeys these conditions.
We can now write eq.\refb{expon} in terms of
\be \label{edefdef}  
T\equiv CS=SC, \qquad M^{rs} \equiv
CV^{rs}\, ,
\ee
and because of \refb{e21aa1}, \refb{econstraint}
we can manipulate the equation as if $T$ and
$M^{rs}$ are numbers rather than infinite dimensional matrices.
We first multiply \refb{expon} by $C$ and write it as: 
\be \label{exponp}
T = X + (M^{12}\,,  M^{21}) (1 - \Sigma\VV)^{-1} 
\pmatrix{TM^{21} \cr TM^{12}}\, ,
\ee
where  
\be \label{e20}
X = M^{11} = C V^{11}\, .
\ee
We then note that since the submatrices commute:
\ben
\label{auxil}
(1 - \Sigma\VV)^{-1}  
&=& \pmatrix{1-TX & -TM^{12} \cr -TM^{21} & 1-TX}^{-1}
\nonumber\\
&=&  ((1-TX)^2 - T^2 M^{12} M^{21})^{-1} \pmatrix{1-TX & TM^{12} \cr
TM^{21} &
1-TX}\, .
\een
Finally, we record that
\be
\label{auxdet}
\hbox{det} (1 - \Sigma\VV) = \hbox{det} ( 1 - 2 TX + T^2 X )\,,
\ee
where use was made the first  equation in \refb{e21aa2} reading
$M^{12} M^{21} = X^2 - X$.

It is now a simple matter to substitute \refb{auxil} into \refb{exponp}
and expand out eliminating all reference 
of $M^{12}$ and $M^{21}$ in favor
of $X$ by use of eqs.\refb{e21aa2}. The result is the condition:  
\be \label{eqfort}
(T-1) ( X T^2 - (1+X) T + X) = 0\, .
\ee
This gives the solution for $S$:\footnote{Of the two other solutions,
$T=1$ gives the identity state $|\II_m\rangle$, whereas the third
solution
has diverging eigenvalues and hence is badly behaved. \label{fo1}}
\be \label{e19}
S = CT, \qquad T= {1\over 2X} (1 + X - \sqrt{(1+3 X)(1 - X)} )\, .
\ee
We can now verify that $S$ obtained this way satisfies
equations \refb{tinv} and \refb{econstraint}.
Indeed, since $CS$ is a function of
$X$, and since
$X(\equiv CV^{11})$   
commutes with $CV^{rs}$, $CS$ also commutes with $CV^{rs}$. Furthermore,
since $V^{11}$ is twist invariant, so is $X$. It then
follows that the inverse
of $X$ and any polynomial in $X$ are twist invariant.  Therefore $T$ is
twist invariant, and, as desired,  $S$ is twist invariant.

\medskip
Demanding that the
normalization factors in
$|\Psi_m\rangle$ and
$|\Psi_m*\Psi_m\rangle$ match gives
\be \label{e21}
\NN = \hbox{det}(1 - \Sigma\VV)^{1/2} = (\det(1 -X) \det(1 + T))^{1/2}
\, ,
\ee
where we have used eqn.\refb{auxdet} and 
simplified it further using \refb{eqfort}. 
Thus the solution is given by
\be \label{e21a}
|\Psi_m\rangle = \{\det(1 - X)^{1/2} \det(1+T)^{1/2} \}^{26}
\exp(-{1\over 2}\, \eta_{\mu\nu} \sum_{m,n\geq 1}S_{mn}\, 
a^{\mu\dagger}_m
a^{\nu\dagger}_n) |0\rangle\, .
\ee
This is the matter part of the state found
in ref.\cite{0008252} (referred to as $|\TT_m\rangle$ in the
introduction) after
suitable correction to the normalization factor.
{}From eq.\refb{e6a} we see that the value of the action
associated with this solution 
has
the form:
\be \label{e22}
\SS|_{\Psi} = 
K\, \NN^{52} \,\langle 0| \exp(-{1\over 2} \eta_{\mu'\nu'} 
\hskip-5pt\sum_{m',n' \geq 1}S_{m'n'}  
a^{\mu'}_{m'}
a^{\nu'}_{n'})
\exp(-{1\over 2} \eta_{\mu\nu}\hskip-5pt \sum_{m,n\geq1} 
S_{mn} a^{\mu\dagger}_m
a^{\nu\dagger}_n) |0\rangle \, .
\ee
By evaluating the matrix element using eq.\refb{egen},
and using the
normalization:
\be \label{e23}
\langle 0| 0\rangle = \delta^{(26)}(0) = {V^{(26)}\over (2\pi)^{26}}\, ,
\ee
where $V^{(26)}$ is the volume of the 26-dimensional space-time,
we get the value of the action to be
\be \label{e24}
\SS|_{\Psi} =  
K\, {V^{(26)}\over (2\pi)^{26}} \, \NN^{52} \{\det(1 -
S^2)^{-1/2}\}^{26}
=
K\,
{V^{(26)}\over (2\pi)^{26}} \, \{\det(1 - X)^{3/4} \det(1 +
3 X)^{1/4}\}^{26}\, .
\ee
In arriving at the right hand side of eq.\refb{e24} we have made use of
eqs.\refb{e19} and \refb{e21}.
Thus the tension of the D25-brane is given by
\be \label{e25}
\TT_{25} = K\, {1\over (2\pi)^{26}}\, \{\det(1 - X)^{3/4} \det(1 +
3 X)^{1/4}\}^{26}\, .
\ee

\subsection{Identification with the sliver 
state} \label{s3}  

In \cite{0006240} a family of surface states was constructed
corresponding to once punctured disks with a special
kind of local coordinates.  They were called wedge states because
the half-disk representing the local coordinates could be viewed
as a wedge of the full unit disk. The puncture was on the boundary
and the wedge has an angle $360^\circ/n$ at the origin, where $n$
is an integer. 
A complete 
description of the state $|n\rangle$ is provided by the fact
that for any state $|\phi\rangle$,\footnote{For brevity,
we have modified the notation of \cite{0006240}. The states
$|{360^\circ\over n}\rangle$ are now simply called $|n\rangle$. Also we
have included an extra scaling by $n/2$ 
in the definition \refb{emapn} of
the
conformal map $f_n$ compared to ref.\cite{0006240}. This does not affect
the definition of $|n\rangle$ due to SL(2,R) invariance of the
correlation functions.} 
\be \label{edefw}
\langle n | \phi\rangle = \langle f_n\circ
\phi(0)\rangle\, ,
\ee
where $f_n\circ \phi(z)$ denotes the conformal transform
of
$\phi(z)$ by the map
\be \label{emapn}
f_n(z) = {n\over 2} \tan\Big({2\over n} \tan^{-1}(z)\Big)\, .
\ee
In the $n\to\infty$ limit this reduces to
\be \label{emap}
f(z) \equiv f_\infty(z) = \tan^{-1}(z)\, .
\ee
It was found in \cite{0006240} that the 
states $|n\rangle$ can be written
in terms of the full Virasoro operators as:
\be
\label{nwedge}
\Bigl|\,{n} \Bigr\rangle = \exp
\Bigl( -{n^2-4\over 3n^2}\, L_{-2} + {n^4 -16\over 30n^4}\, L_{-4} -
{(n^2-4) (176+ 128 n^2 + 11 n^4)\over 1890 n^6} \,L_{-6}
 + \cdots \Bigr) | 0\rangle \,.
\ee
For $n=1$ the state reduces to the identity string field:
$|n=1\rangle = |\II\rangle$.
For $n=2$ we get the vacuum:
$|n=2\rangle = |0\rangle$.
For $n \to \infty$, which
corresponds to a vanishingly thin  wedge state, 
and will be called the sliver state $|\WW\rangle$, 
we find a smooth limit
\be \label{smoothlimit}  
|\WW\rangle \equiv |\infty\rangle = \exp
\Bigl( -{1\over 3} L_{-2} 
+ {1\over 30} L_{-4} - {11\over 1890} L_{-6} +
{34\over 467775} L_{-8}
 + \cdots \Bigr) | 0\rangle \,.
\ee
It was also shown in \cite{0006240} that  
\be
| n\rangle * | m \rangle = | n+ m -1\rangle \,.
\ee
Thus the
state $|\WW\rangle$ has the property that $|n\rangle * |
\WW\rangle = |\WW\rangle$ for any $n\geq 1$.  In particular,
$|\WW\rangle$ squares to itself.
Some properties
of this state
have been discussed in appendix \ref{a3}.

\begin{table}
\begin{center}\def\st{\vrule height 3ex width 0ex}
\begin{tabular}{|l|l|l|l|l|l|l|} \hline
$L$ & $S_{11}$ & $S_{22}$ & $S_{13}$  & $S_{33}$ & $S_{24}$ & $S_{44}$
\st\\[1ex]
\hline
\hline
20  & 0.2888 &  $-$0.0627 & $-$0.1263 & 0.0706 & 0.0440 & $-$0.0347
\st\\[1ex]
\hline
40 & 0.2970 & $-$0.0638 & $-$0.1316 & 0.0740 & 0.0451 & $-$0.0358
\st\\[1ex]
\hline
80 & 0.3033 & $-$0.0647 & $-$0.1356 & 0.0766 & 0.0459 & $-$0.0365
\st\\[1ex]
\hline
160 & 0.3081 &  $-$0.0652 & $-$0.1387 & 0.0785 & 0.0465 & $-$0.0371
\st\\[1ex]
\hline
$\infty$ & 0.3419 & $-$0.0665 & $-$0.1588 &
0.0905  & 0.0476 & $-$0.0382
\st\\[1ex]
\hline
\end{tabular}
\end{center}
\caption{ Numerical results for the elements of the matrix
$S$.
We compute $S$ by restricting the indices $m,n$ of $V^{11}_{mn}$ and
$C_{mn}$ to be $\le L$ so that $V^{11}$ and $C$ are $L\times L$
matrices,
and then using eq.\refb{e19}.
The last row
shows the interpolation of the various results
to $L=\infty$, obtained via a fitting function of the
form $a_0 + a_1/\ln(L) + a_2/(\ln(L))^2+
a_3/(\ln(L))^3 $.} \label{t2}
\end{table}

Given the split $L= L^m + L^{g}$
of the Virasoro operators into
commuting Virasoro subalgebras, the state $|\WW\rangle$ 
can be written in factorized
form: an exponential of matter Virasoros, and an identical exponential
of ghost Virasoros:
\be
|\WW\rangle = |\WW_m\rangle \otimes |
\WW_{g}\rangle \, .
\ee
In particular, it follows from \refb{smoothlimit} that
\be \label{eexpwm} 
|\WW_m\rangle = \wh\NN^{26} \exp   
\Bigl(-{1\over 3} L_{-2}^m + {1\over 30} L_{-4}^m - {11\over 1890}
L_{-6}^m +
 \cdots \Bigr) | 0\rangle \, , 
\ee
where $\wh\NN$ is a normalization factor to be fixed shortly.  
The property $|\WW\rangle * |\WW\rangle = |\WW\rangle$ implies
that
\be \label{elambda}
|\WW_m\rangle *^m |\WW_m\rangle
= \lambda \wh\NN^{52} |\WW_m\rangle  
\ee
where the constant $\lambda$ could possibly be vanishingly small or
infinite. We shall choose $\wh\NN$ such that 
$\lambda \wh\NN^{52}=1$,  
so that $|\WW_m\rangle$ squares to itself.

In order to show that this sliver state 
is the matter state identified in the previous subsection
for the D25-brane, we must compare \refb{eexpwm} with
\be \label{ecomp}  
|\Psi_m\rangle \,  =  
\NN^{26} \exp \Bigl( - {1\over 2}   
\eta_{\mu\nu} a^{\mu\dagger} 
\cdot S
\cdot a^{\nu\dagger} \Bigr) |0\rangle\, ,  
\ee
where $S$ is the matrix calculated in the previous
subsection.
Since the Virasoro operators contain both
positively and negatively moded oscillators a comparison requires
expansion.
While we have done this as a check, 
the techniques of ref.\cite{LPP} enable one to use the local
coordinate 
\refb{emap} to give a direct oscillator 
construction
of the sliver state 
\be \label{ehat}  
|\WW_m\rangle = \wh \NN^{26} \exp(-   
{1\over 2}\eta_{\mu\nu} a^{\mu\dagger} 
\cdot \wh S\cdot
a^{\nu\dagger})\, ,
\ee
where,
\be \label{es1}
\wh S_{mn} = -{1\over \sqrt{mn}} \,  \ointop
{d w\over
2\pi i}\, \ointop {dz\over 2\pi i} {1 \over z^n w^m (1 + z^2) (1+w^2)
(\tan^{-1}(z) -
\tan^{-1}(w))^2}\, .
\ee
$\ointop$ denotes a contour integration around the origin. 
As required by twist invariance, $\wh S_{mn}$ vanishes 
when $m+n$ is odd.  
Explicit computations
give:
\ben \label{es2}
&&\wh S_{11} = {1\over 3} \simeq .3333\, , \quad\,\,\,\,\,
\wh S_{22} = -{1\over 15} \simeq -.0667 \, , \quad
\wh S_{13}= -{4 \over 15 \sqrt{3}} \simeq -.1540 \, , \nonumber \\
&&\wh S_{33}= {83 \over 945} \simeq .0878 \, , \quad
\wh S_{24} = {32 \sqrt{2} \over 945} \simeq .0479 \, ,\quad
\wh S_{44} = - {109 \over 2835} \simeq -.0384\, .
\een
On the other hand a level expansion computation for $S_{mn}$, together
with a fit, has been shown
in table \ref{t2}. The data shows rather remarkable
agreement between $S_{mn}$ and $\wh S_{mn}$. The errors are of the order
of 3\%. 
Once $S_{mn}$ and $\wh S_{mn}$ agree, the normalization 
factors must agree as well, since both states square to themselves under
$*^m$-product.  
This is convincing evidence 
that the matter part of the state representing the
D25-brane solution is identical to the matter
part $|\WW_m\rangle$
of the sliver state up to an overall normalization.

\sectiono{Construction of the lump solutions} \label{s3a}

In this section we shall discuss the 
construction of 
lump solutions of
eq.\refb{e5}
representing lower dimensional D-branes. We are able to  
give these solutions in closed form and to express the
ratio of tensions of branes of different dimensions in
terms of determinants of infinite dimensional 
matrices. Numerical
calculation of these ratios in the level expansion gives
remarkable agreement with the expected values.

\subsection{Lump solutions and their tensions} \label{ss31}

We begin by noting that  
the solution \refb{e21a} representing the D25 brane  
has the form of a 
product over 26 factors, each involving the oscillators associated
with a given direction. This suggests that in order to construct a
solution of codimension $k$ representing a D-$(25-k)$-brane, we need to
replace $k$ of
the factors associated with directions transverse to the D-brane by a
different set of solutions, but the factors associated with directions
tangential to the D-brane remains the same. (This is precisely what
happens
in the case of $p$-adic string theory 
\cite{Brekke,0003278,0102071}, background independent open string field
theory \cite{9208027,9210065,9303067,9303143,9311177,0008231,
0009103,0009148,0009191}, as well as non-commutative
solitons \cite{0003160,0005006,0005031}.) 
A procedure for constructing such space(-time) dependent solutions
was given in
ref.\cite{0008252}. Suppose we are interested in a D-$(25-k)$ brane
solution. Let us denote by $x^\bmu$ ($0\le \bmu\le (25-k)$) the
directions tangential to the brane and by $x^\alpha$ ($(26-k)\le
\alpha\le 25$) the
directions transverse to the brane. We now use the representation of the
vertex in the zero mode oscillator basis for the directions $x^\alpha$,
as given in appendix \ref{a4}.
For this we define, for each string,
\be \label{e266a}
a^\alpha_0 = {1\over 2}\, \sqrt b\,  \hat p^\alpha  - {1\over \sqrt b}
\,
i
\hat x^\alpha ,
\qquad a^{\alpha\dagger}_0={1\over 2}\, \sqrt b \, \hat p^\alpha +
{1\over
\sqrt
b} \,
i \hat x^\alpha \, ,
\ee
where
$b$ is an arbitrary constant, and $\hat x^\alpha$ and $\hat p^\alpha$
are
the zero mode coordinate and momentum operators associated with the
direction $x^\alpha$. We also denote by
$|\Omega_b\rangle$ the normalized state
which is annihilated by all the annihilation operators $a_0^\alpha$, and
by $|\Omega_b\rangle_{123}$ the direct product of the vacuum
$|\Omega_b\rangle$ for each of the three strings. As shown in
appendix \ref{a4} (eq.\refb{e288}), the  vertex $|V_3  
\rangle$
defined in eq.\refb{e9} 
can be rewritten in this new basis as:
\ben \label{e28}
|V_3 \rangle &=&
\int d^{26-k}p_{(1)} d^{26-k}p_{(2)} d^{26-k}p_{(3)}
\delta^{(26-k)}(p_{(1)} + p_{(2)} +
p_{(3)}) \nonumber \\
&& \hskip-13pt\exp \Bigl(-{1\over 2}
\sum_{r, s\atop m, n \ge 1} 
\eta_{\bmu\bnu} a^{(r)\bmu\dagger}_m
V^{rs}_{mn} a_n^{(s)\bnu\dagger}
- \sum_{r,s\atop n \ge 1}  
 \eta_{\bmu\bnu} p_{(r)}^{\bmu} V^{rs}_{0n}
a_n^{(s)\bnu\dagger}
-{1\over 2}\sum_r\eta_{\bmu\bnu} p_{(r)}^{\bmu}
V^{rr}_{00}
p_{(r)}^{\bnu} \Bigr) |0, p\rangle_{123} \nonumber \\
&& \otimes \, \bigg( {\sqrt 3 \over (2\pi b^3)^{1/4}}
(V^{rr}_{00}+{b\over 2})\bigg)^{-k}
\exp \Bigl(-{1\over 2}
\sum_{r, s\atop m, n \ge 0} 
a^{(r)\alpha\dagger}_m V^{\prime rs}_{mn}
a_n^{(s)\alpha\dagger} \Bigr)
|\Omega_b\rangle_{123} \, . 
\een
In this expression the sums over $\bmu, \bnu$
run from 0 to $(25-k)$, and sum
over $\alpha$ runs from $(26-k)$ to 25. Note that in the last line the
sums
over $m$, $n$ run over 0, 1, 2 $\ldots$. The coefficients
$V^{\prime r s}_{mn}$ have been given in terms of $V^{rs}_{mn}$ in
eq.\refb{e299}.

In Appendix \ref{a4} it is shown that
$V^{\prime rs}$, regarded as matrices with indices running from 0 to
$\infty$,
satisfy (see \refb{e30} and \refb{e311})
\be \label{e30p}
V^{\prime\, r\, s} = {1\over 3} ( C' + \omega^{s-r} U' + \omega^{r-s}\wb
U')\,,
\ee
where we have dropped the explicit $b$ dependence from the notation,
$C'_{mn}=(-1)^m\delta_{mn}$
with indices $m,n$ now running from 0 to
$\infty$, and
$U'$,
$\wb U' \equiv U'^*$ viewed as  matrices with $m,n\geq 0$
satisfy the relations:
\be \label{e311p}
\wb U' = C' U' C', \qquad U^{\prime 2} = \wb U^{\prime 2} = 1, \qquad
U^{\prime \dagger} = U'\, .
\ee
We note now the complete analogy with equations \refb{e11p} and
\refb{e133p} \cite{0008252}.
It follows also that the $V'$ matrices,
together
with $C'$ will satisfy equations exactly analogous to
\refb{e21aa1}, \refb{e21aa2}.
Thus
we can construct a
solution of the equations of motion \refb{e5} in an identical manner
with the
unprimed quantities replaced by the primed quantities. Taking into
account
the extra normalization factor appearing in the last line of
eq.\refb{e28}, we get the following form of the solution of
eq.\refb{e5}:
\ben \label{e32}
|\Psi_m'\rangle &=& \{\det(1 - X)^{1/2} \det(1+T)^{1/2} \}^{26-k}
\exp(-{1\over 2} \eta_{\bmu\bnu} \sum_{m,n\ge 1} S_{mn}
a^{\bmu\dagger}_m
a^{\bnu\dagger}_n) |0\rangle \nonumber \\
&& \otimes \bigg( {\sqrt 3 \over (2\pi b^3)^{1/4}}
(V^{rr}_{00}+{b\over 2})\bigg)^{k}
\{\det(1 - X')^{1/2}\det(1+T')^{1/2} \}^{k} \nonumber \\
&&\qquad \exp \Bigl(-{1\over 2}
\sum_{m,n\ge 0} S'_{mn}
a^{\alpha\dagger}_m
a^{\alpha\dagger}_n\Bigr) |\Omega_b\rangle\, ,
\een
where
\be \label{e33}
S' = C' T'\, , \qquad T' = {1\over 2X'} (1 + X' - \sqrt{(1+3 X')(1 -
X')}
)\, ,
\ee
\be \label{e34}
X' = C' V^{\prime 11}\, .
\ee
Using eq.\refb{e6a} we can calculate the value of the action associated
with this solution. It is given by an equation analogous to \refb{e24}:
\ben \label{e35}
\SS_{\Psi'} &=& 
K\,
{V^{(26-k)}\over (2\pi)^{26-k}} \, \{\det(1 - X)^{3/4} \det(1 +
3 X)^{1/4}\}^{26-k} \nonumber \\
&& \times \bigg( { 3 \over (2\pi b^3)^{1/2}}
(V^{rr}_{00}+{b\over 2})^2\bigg)^{k} \{\det(1 - X')^{3/4} \det(1 +
3 X')^{1/4}\}^k \, ,
\een
where $V^{(26-k)}$ is the D-$(25-k)$-brane world-volume.
This gives the tension of the D-$(25-k)$-brane to be
\ben \label{e36}
\TT_{25-k} &=& K\, {1\over (2\pi)^{26-k}}\, \{\det(1 - X)^{3/4} \det(1 +
3 X)^{1/4}\}^{26-k} \nonumber \\
&& \times \bigg( { 3 \over (2\pi b^3)^{1/2}}
(V^{rr}_{00}+{b\over 2})^2\bigg)^{k} \{\det(1 - X')^{3/4} \det(1 +
3 X')^{1/4}\}^k \, . 
\een
Clearly for $k=0$ this agrees with \refb{e25}. From eq.\refb{e36} we get
\be \label{e37}
{\TT_{24-k} \over 2\pi \TT_{25-k}} = { 3 \over \sqrt{2\pi
b^3}} \,
\Bigl(V^{rr}_{00}+{b\over 2}\Bigr)^2\, {\{\det(1 - X')^{3/4} \det(1 +
3 X')^{1/4}\} \over \{\det(1 - X)^{3/4} \det(1 +
3 X)^{1/4}\}}\, .
\ee

\begin{table}
\begin{center}\def\st{\vrule height 3ex width 0ex}
\begin{tabular}{|l|l|l|l|l|l|l|} \hline
L & ${\TT_{k} \over 2\pi \TT_{k+1}}$ for & ${\TT_{k} \over 2\pi
\TT_{k+1}}$
for & ${\TT_{k} \over 2\pi \TT_{
k+1}}$ for & ${\TT_{k} \over 2\pi \TT_{
k+1}}$ for & ${\TT_{k} \over 2\pi \TT_{
k+1}}$ for & $(2\pi)
\Big({\TT_{25}\over K}\Big)^{1/26}$
\st\\[1ex]
& $b=.5$ & $b=1$ & $b=2$ & $b=4$ & $b=8$ &
\st\\[1ex]
\hline
\hline
0 & 1.38117  &1.08677 & .96195 & .94748 & 1.00992 &  1 \st\\[1ex]
\hline
50 & .94968 & .95490 & .96518 & .98485 & 1.02101 & .69705 \st\\[1ex]
\hline
100 & .95488 & .95959 & .96884 & .98659 & 1.01946 & .64875 \st\\[1ex]
\hline
200 & .95906 & .96333 & .97172 & .98789 & 1.01802 & .60244 \st\\[1ex]
\hline
400 & .96250 & .96640 & .97408 & .98891 & 1.01672 & .55839 \st\\[1ex]
\hline
800 & .96539 & .96898 & .97606 & .98976 & 1.01558 & .51718 \st\\[1ex]
\hline
1600 & .96785 & .97118 & .97775 & .99048 & 1.01456  & .48082\st\\[1ex]
\hline
3200 & .97007 & .97316 & .97927 & .99113 & 1.01361 & .44664 \st\\[1ex]
\hline
$\infty$ & 1.00042 & 1.00061 & 1.00063 & 1.00031 & .99939 & -.12638
\st\\[1ex]
\hline
\end{tabular}
\end{center}
\caption{ Numerical results for the ratio
${\TT_{k} \over 2\pi \TT_{k+1}}$.
The first column shows the level up to which we calculate the
matrices $V^{rs}_{mn}$ and $V^{\prime rs}_{mn}$. The second to sixth
column shows the ratio
${\TT_{k} \over 2\pi \TT_{k+1}}$
for different values of the parameter $b$. The last column gives
$(2\pi)^{26}\TT_{25}/K$. The last row shows the interpolation of the
various results
to $L=\infty$, obtained via a fitting function of the
form $a_0 + a_1/\ln(L) + a_2/(\ln(L))^2 + a_3/(\ln(L))^3$.} \label{t1}
\end{table}

This ratio can be calculated if we restrict $m,n$ to be below a
given level $L$, so that $X=CV^{11}$ is an $L\times L$ matrix and
$X'=C'V^{\prime 11}$
is an $(L+1)\times (L+1)$ matrix.
The values of $V^{11}$ and $V^{\prime 11}$ can be found from
eqs.\refb{ea3} and \refb{e299}.  
In particular for $L=0$ only the matrix $X'$ contributes. {}From
eq.\refb{e34}, \refb{e299} and \refb{ea3} we get
\be \label{elevel0} 
V^{11}_{00} = \ln{27\over 16}, \qquad
X'_{00} = V^{\prime 11}_{00} = 1 -{2\over 3}\, {b\over \ln(27/16) +
{b\over
2}}\, .
\ee
Thus in the $L=0$ approximation, the ratio \refb{e37} is given by
\be \label{elevel02}
{ 3 \over \sqrt{2\pi}} \, \Big({2\over 3} \Big)^{3/4} (4
\ln(27/16)
)^{1/4}
\Big(\ln(27/16) +{b\over 2}\Big)\, b^{-3/4}\, .
\ee
For larger values of $L$ the ratio is calculated numerically.
The results
of
the numerical analysis are given in table \ref{t1}. As seen from this
table, the ratio $\TT_k / (2\pi \TT_{k-1})$ 
approaches 1 as $L\to\infty$
for all $b$. This is exactly what is expected if the lump solutions
discussed here describe lower dimensional D-branes. It is also seen from
the table that $\TT_{25}/K$ extrapolated to $L=\infty$ gives a
negative number. We take this as an evidence that it approaches 0 as
$L\to\infty$. This
indicates that the matter component $|\Psi_m\rangle$ of the string field
has zero norm.
We expect that this will be compensated by the ghost sector
contribution $K$, so that the contribution
to the action from the full string
field approaches a finite limit as we take the level of approximation
$L$
to $\infty$.

As in the previous section (footnote \ref{fo1}), one can construct two
other
solutions to eq.\refb{e5}. For one of them $T'$ is the inverse of the
solution for $T'$ given in eq.\refb{e33}. For this solution the
eigenvalues
of $T'$ diverge and so the state is not well behaved. The other solution
corresponds to $T'_{mn}=\delta_{mn}$, 
{\it i.e.} $S'_{mn}=C'_{mn}$. Using
eq.\refb{e27} for $p^\alpha=0$ 
to go from oscillator
basis to the momentum basis, one can
easily verify that this again is the identity string state
$|\II_m\rangle$. Thus we do not get a new solution.

\subsection{$b$-dependence of the solution} \label{ss32}

The analysis of the last section generates a one-parameter family of
lump
solutions characterized by the parameter $b$.\footnote{Actually for a
codimension $k$ lump we have a $k$ parameter family of solutions since
we
can choose different parameters $b$ corresponding to different
directions.}
Thus we are now faced with an embarrassment of riches, $-$ for these
solutions to have the interpretation as D-branes there should be a
unique solution (up the possibility of translating the solution in the
transverse
direction) and not a family of solutions. There are two possibilities
that come to mind.
\begin{enumerate}
\item Although in the oscillator basis the solution seems to depend on
$b$, the relationship between the oscillator and the momentum basis is
$b$-dependent, and when we rewrite the solution in the momentum basis
it is actually $b$-independent.
\item Even after rewriting the solution in the momentum basis it is
$b$-dependent, but the
solutions for different values of $b$  
are related to each other by gauge transformation.
\end{enumerate}

We shall begin by exploring the first possibility. In order to get basis
independent information about the lump solution, we can calculate its
inner product
with states in the momentum basis. Let us, for example, consider the
inner
product $\langle \{p^\alpha\}|\Psi_m'\rangle$. Using eqs.\refb{e27},
\refb{e32} and \refb{egen} we get
\be \label{en1}
\langle \{p^\alpha\}|\Psi_m'\rangle \propto \exp\Big(-{p^2\over
2}({b\over
2}
+
b{S'_{00}\over 1 - S'_{00}}\big)\Big)\, ,
\qquad p^2 = p^\alpha p^\alpha\, .
\ee
The numerical results for the values of $S'_{00}$ for different values
of
$b$ are shown in table \ref{t11}, and the values of $b/2 + b S'_{00}/ (1
- S'_{00})$ have been shown in table \ref{t12}. {}From this we see
clearly
that $\langle \{p^\alpha\}|\Psi_m'\rangle$ is not independent of $b$.

\begin{table}
\begin{center}\def\st{\vrule height 3ex width 0ex}
\begin{tabular}{|l|l|l|l|l|l|} \hline
L & $S'_{00}$ for & $S'_{00}$
for & $S'_{00}$ for & $S'_{00}$ for & $S'_{00}$ for \st\\[1ex]
& $b=1$ & $b=2$ & $b=4$ & $b=6$ & $b=8$   
\st\\[1ex]
\hline
\hline
20 & .25980 & .08001 & $-$.10718 & $-$.21475 & $-$.28836 \st\\[1ex]
\hline
40 & .25796 & .07692 & $-$.11183 & $-$.22034 & $-$.29458 \st\\[1ex]
\hline
60 & .25711 & .07548 & $-$.11403 & $-$.22302 & $-$.29758 \st\\[1ex]
\hline
80 & .25659 & .07460 & $-$.11539 & $-$.22469 & $-$.29947 \st\\[1ex]
\hline
100 & .25623  & .07398  & $-$.11636  & $-$.22588  & $-$.30081 \st\\[1ex]
\hline
120 & .25596  & .07352  & $-$.11709  & $-$.22678  & $-$.30184 \st\\[1ex]
\hline
$\infty$ & .2497  & .0619  & $-$.1372  & $-$.2534  & $-$.3333 \st\\[1ex]
\hline
\end{tabular}
\end{center}
\caption{ Numerical results for $S'_{00}$. The first column shows the
level up to which we calculate the
matrices $V^{rs}_{mn}$ and $V^{\prime rs}_{mn}$. The second to sixth
column shows $S'_{00}$ for different values of the parameter $b$. The
last row shows the interpolation of the
various results
to $L=\infty$, obtained via a fitting function of the
form $a_0 + a_1/\ln(L) + a_2/(\ln(L))^2 +
a_3/(\ln(L))^3$.} \label{t11}
\end{table}
\begin{table}
\begin{center}\def\st{\vrule height 3ex width 0ex}
\begin{tabular}{|l|l|l|l|l|l|} \hline
$b$ & $1$ & $2$ & $4$ & $6$ & $8$
\st\\[1ex]
\hline
$b/2 + b S'_{00}/ (1 - S'_{00})$ &
.833 & 1.132 & 1.517  & 1.787 & 2.000 \st\\[1ex]
\hline
\hline
\end{tabular}
\end{center}
\caption{ Numerical results for $b/2 + b S'_{00}/ (1 - S'_{00})$
for different values of $b$. We use the results for $S'_{00}$
given in the last row of table \ref{t11}.} \label{t12}
\end{table}

This brings us to the second possibility: could the different solutions
be
related by gauge transformation? Since the different solutions have the
same ghost component, such a gauge transformation must be of a special
kind that changes the matter part but not the ghost part. So the
question
is: are such gauge transformations possible?  If we choose the gauge
transformation parameter
$|\Lambda\rangle$
to be of the form $|\Lambda_g\rangle \otimes |\Lambda_m\rangle$, then,
under a gauge transformation
\be \label{en2}
\delta (|\Psi_g\rangle\otimes |\Psi_m\rangle)
= \QQ |\Lambda_g\rangle\otimes |\Lambda_m\rangle + |\Psi_g *^g
\Lambda_g\rangle\otimes
|\Psi_m *^m \Lambda_m\rangle
- |\Lambda_g *^g \Psi_g\rangle\otimes |\Lambda_m *^m \Psi_m\rangle\, .
\ee
Now suppose $|\Lambda_g\rangle$ is such that
\be \label{en3}
\QQ |\Lambda_g\rangle = 0, \qquad |\Lambda_g *^g \Psi_g\rangle = |\Psi_g
*^g
\Lambda_g\rangle = |\Psi_g\rangle\, .
\ee
In that case \refb{en2} can be written as
\be \label{en4}
\delta (|\Psi_g\rangle\otimes |\Psi_m\rangle)
= |\Psi_g\rangle\otimes (|\Psi_m *^m \Lambda_m\rangle - |\Lambda_m *^m
\Psi_m\rangle)\, .
\ee
Thus effectively the gauge transformation induces a transformation on
the
matter part of the solution without any transformation on the
ghost sector. It is our guess that solutions with different values of
$b$
are related by gauge transformations of this kind. Although
we do not
have
a complete proof of this, a partial analysis of this problem has been
carried out in appendix \ref{a2}.
If this is indeed true,
then this will imply that the width of the solution in the position
space,
given by $\sqrt{b/2 + b S'_{00}/ (1 - S'_{00})}$, is a gauge dependent
quantity.

Without having detailed knowledge of the operator $\QQ$ we cannot know
whether there is some ghost number zero state $|\Lambda_g\rangle$ in the
ghost sector satisfying eq.\refb{en3}. Note however that if $\QQ$
annihilates the identity $|\II\rangle$ of the $*$ product then taking
$|\Lambda_g\rangle= |\II_g\rangle$, where $|\II_g\rangle$ denotes the
component of $|\II\rangle$  in the ghost sector,
automatically
satisfies
eq.\refb{en3}. On the other hand since eq.\refb{en3} needs to be
satisfied
only for a special $|\Psi_g\rangle$ which represents D-brane solutions,
there may be other $|\Lambda_g\rangle$ satisfying these equation.

Note that even if we did not discover the existence of multiple
solutions
labeled by different values of $b$, we would still have an
embarrassment
of riches if there were no $|\Lambda_g\rangle$ satisfying eq.\refb{en3}.
This is due to the fact that given any solution of
eq.\refb{e5}, we can generate other solutions by deforming
$|\Psi_m\rangle$ as follows:
\be \label{en6}
\delta|\Psi_m\rangle
= |\Psi_m *^m \Lambda_m\rangle - |\Lambda_m *^m
\Psi_m\rangle\, .
\ee
In order to make sense of these solutions, we must show that when we
combine them with $|\Psi_g\rangle$ to construct solutions of the full
string field theory equations of motion, they
are related by gauge transformations. The postulate of existence of a
$|\Lambda_g\rangle$ satisfying eq.\refb{en3} makes this possible.

\sectiono{Open questions} \label{s5}

Clearly many questions remain unanswered. In this concluding section we
shall try to make a list of questions which we hope will be answered in
the near future.

\begin{enumerate}

\item The most pressing question at this time seems to be understanding
the ghost sector of the solution. We expect that a proper analysis of
the
ghost sector will not only lead to the solution, but will also fix
uniquely (up to the field redefinition ambiguity) the form of the
kinetic operator $\QQ$.
Since the matter part of the D25-brane solution is given by the matter
part $ \WW_m $ of the sliver state, 
our guess is that
the
full solution is given by a ghost number one operator
built purely out of ghosts
acting on the full sliver state  
$\WW $.

\item

For the D-branes of dimension $<25$, we have found families of candidate
solutions labeled by the parameter $b$. Since a physical D-brane does
not admit continuous deformations other than the translational motion
transverse to the brane, we need to show that these additional
deformations are gauge artifacts. We have given some arguments to this
effect in appendix \ref{a2}, but a complete proof is lacking.

\item Although we have shown that the ratios of tensions of our
solutions
agree very well with the expected answer, in order to establish
conclusively that these solutions describe D-branes, we need to analyze
the fluctuations of the string field around these solutions and show
that
the spectrum and interaction of these fluctuations agree with those of
conventional open strings living on the D-brane. Clearly, the knowledge
of
$\QQ$ is crucial for this study.

\item

Although the matter parts of our solutions are given in analytic form,
calculation of the ratios of tensions of these solutions, involving
computation of determinants of infinite dimensional matrices, was done
numerically. It will be nice to have an analytic expression for this
ratio.

\item If our solution really describes D-branes, then we expect that
there
should be static multiple lump solutions representing multiple D-branes.
One should be able to construct such solutions in our string field
theory.
It is natural to assume that these multi-lump solutions will also have
factorized form, with the ghost part being described by the same
universal
state as the single lump solutions. Thus this analysis can  be carried
out
without a detailed knowledge of $\QQ$.

\item
The procedure that we have followed to construct  
lower dimensional branes from the D25 brane solution
bears a suggestive 
formal similarity with the solution
generating techniques which have appeared
recently in studies of non-commutative solitons
and have been conjectured to be relevant to string field theory
\cite{0003160,0010034,0010060}. 
In that context, new space-time dependent solutions
are obtained by 
acting on a translational invariant solution
with ``non-unitary isometries'', like the ``shift operator''
in an infinite dimensional Hilbert 
space.
The construction of the matrix $T'$ from $T$ is quite
reminiscent of some sort of shift operation.
It would be interesting to investigate this connection precisely.

\end{enumerate}

\medskip
\bigskip

\noindent{\bf Acknowledgements}:
We would like to thank D. Gaiotto, J. Minahan, N. Moeller, M.~Schnabl, 
W.~Taylor and E. Witten for useful discussions.
The work of L.R. was supported in part
by Princeton University
``Dicke Fellowship'' and by NSF grant 9802484.
The work of  B.Z. was supported in part
by DOE contract \#DE-FC02-94ER40818.

\appendix

\sectiono{The coefficients $V^{rs}_{mn}$} \label{a1}

In this appendix we give the coefficients $V^{rs}_{mn}$ introduced in
the
text. These results are taken from refs.\cite{gross-jevicki,gj2}.
First we define the coefficients $A_n$ and $B_n$ for $n\ge 0$ through
the
relations:
\be \label{ea1}
\bigg({1 + i x \over 1 - i x}\bigg)^{1/3} = \sum_{n\, even} A_n
x^n
+ i \sum_{n\, odd} A_n x^n\, , \qquad
\bigg({1 + i x \over 1 - i x}\bigg)^{2/3} = \sum_{n\, even} B_n
x^n
+ i \sum_{n\, odd} B_n x^n\, .
\ee
In terms of $A_n$ and $B_n$ we define the coefficients $N^{r, \pm
s}_{mn}$
as follows:
\ben \label{ea2}
N^{r, \pm r}_{nm} &=& {1 \over 3 (n \pm m)} \, (-1)^n (A_n B_m \pm B_n
A_m)
\quad
\hbox{for} \quad m+n \, \, \hbox{even}, \, \, m\ne n\, , \nonumber \\
&=& 0 \quad
\hbox{for} \quad m+n \, \, \hbox{odd}\, , \nonumber \\
N^{r, \pm (r+1)}_{nm} &=& {1 \over 6 (n \pm m)} \, (-1)^{n+1} (A_n B_m
\pm
B_n A_m)
\quad
\hbox{for} \quad m+n \, \, \hbox{even}, \, \, m\ne n\, , \nonumber \\
&=& {1 \over 6 (n \pm m)} \, \sqrt{3}\, (A_n B_m \mp
B_n A_m)
\quad
\hbox{for} \quad m+n \, \, \hbox{odd}\, , \nonumber \\
N^{r, \pm (r-1)}_{nm} &=& {1 \over 6 (n \mp m)} \, (-1)^{n+1} (A_n B_m
\mp
B_n A_m)
\quad
\hbox{for} \quad m+n \, \, \hbox{even}, \, \, m\ne n\, , \nonumber \\
&=& -{1 \over 6 (n \mp m)} \, \sqrt{3}\, (A_n B_m \pm
B_n A_m)
\quad
\hbox{for} \quad m+n \, \, \hbox{odd}\, .
\een
The coefficients $V^{rs}_{mn}$ are then given by
\ben \label{ea3}
V^{rs}_{nm} &=& - \sqrt{mn}\, (N^{r,s}_{nm}+ N^{r, -s}_{nm})\quad
\hbox{for} \quad m\ne n,\, m, n \ne 0\, , \nonumber \\
V^{rr}_{nn} &=& -{1\over 3} [ 2\sum_{k=0}^n (-1)^{n-k} A_k^2 - (-1)^n -
A_n^2 ], \quad \hbox{for} \quad n\ne 0\, , \nonumber \\
V^{r (r+1)}_{nn} &=& V^{r (r+2)}_{nn} = {1\over 2} [(-1)^n -
V^{rr}_{nn}]
\quad \hbox{for} \quad n\ne 0\, ,\nonumber \\
V^{rs}_{0 n}  &=& - \sqrt{2 n}\, (N^{r,s}_{0 n}+ N^{r, -s}_{0 n})\quad
\hbox{for} \quad n \ne 0\, , \nonumber \\
V^{rr}_{00} &=& \ln(27/ 16) \, .
\een
The value of $V^{rr}_{nn}$ quoted above corrects the result for
$N^{rr}_{nn} (\equiv - V^{rr}_{nn}/n)$
quoted in eqn.(1.18) of \cite{gj2}.
In writing down the expressions
for $V^{rs}_{0n}$ and $V^{rr}_{00}$ we
have taken into account the fact that we are using $\alpha'=1$
convention,
as opposed to the $\alpha'=1/2$ convention used in
refs.\cite{gross-jevicki, gj2}.

Finally we would like to point out that our convention for
$|A*^mB\rangle$, defined through eq.\refb{e8}, differs from that in
refs.\cite{gross-jevicki, gj2}. In particular, with the values of
$V^{rs}_{mn}$ give in eq.\refb{ea3}, our $|A*^mB\rangle$ would
correspond to $|B*^mA\rangle$ in ref.\cite{gross-jevicki}. Since
the string field
equation of motion involves $|\Psi*\Psi\rangle$, it is not
affected by this difference in
convention. However, if we want
our convention for $|A*^mB\rangle$ to agree with that of
ref.\cite{gross-jevicki}, we should replace $V^{rs}$ by $V^{sr}$
everywhere in eqs.\refb{ea3}.

\sectiono{Conversion from momentum to oscillator basis} \label{a4} 

We start with the three string vertex in the matter sector as given in
section \refb{ss22}:
\be \label{e99}
|V_3 \rangle = \int d^{26}p_{(1)} d^{26}p_{(2)} d^{26}p_{(3)}
\delta^{(26)}(p_{(1)} + p_{(2)}
+ p_{(3)}) \exp (-  E) |0, p\rangle_{123}
\ee
where
\be
\label{e99p}
E =  {1\over 2} \sum_{{r,s}\atop m, n \ge 1}
 \eta_{\mu\nu} a^{(r)\mu\dagger}_m V^{rs}_{mn}
a_n^{(s)\nu\dagger} +
\sum_{{r,s}\atop n\ge 1}
 \eta_{\mu\nu} p_{(r)}^{\mu} V^{rs}_{0n}
a_n^{(s)\nu\dagger}
+{1\over 2}\sum_r\eta_{\mu\nu} p_{(r)}^{\mu}
V^{rr}_{00} p_{(r)}^{\nu}  \, .
\ee
Note that
using the freedom of redefining $V^{rs}_{00}$ using momentum
conservation,
we have chosen $V^{rs}_{00}$ to be zero for $r\ne s$. Due to the same
reason, a redefinition $V^{rs}_{0n} \to V^{rs}_{0n} + A^s_n$ by some $r$
independent constant $A^s_n$ leaves the vertex unchanged. We shall use
this freedom to choose:
\be \label{evcon}
\sum_r V^{rs}_{0n} = 0\, .
\ee
It can be easily verified that $V^{rs}_{0n}$ 
given in eq.\refb{ea3} satisfy these conditions.

We now pass   
to the oscillator basis for a subset of the space-time coordinates
$x^\alpha$ ($(26-k)\le \alpha\le
25$), by relating the
zero mode operators
$\hat x^\alpha$ and $\hat p^\alpha$ to oscillators $a^\alpha_0$ and
$a^{\alpha\dagger}_0$. For this
one writes:
\be \label{e266}
a^\alpha_0 = {1\over 2}\, \sqrt b\,  \hat p^\alpha  - {1\over \sqrt b}
\,
i
\hat x^\alpha ,
\qquad a^{\alpha\dagger}_0={1\over 2}\, \sqrt b \, \hat p^\alpha +
{1\over
\sqrt
b} \,
i \hat x^\alpha \, ,
\ee
where
$b$ is an arbitrary constant. Then $a^\alpha_0$,
$a^{\alpha\dagger}_0$ satisfy the usual commutation rule
$[a_0^\alpha , a_0^{\beta\dagger} ] = \delta^{\alpha\beta}$ (we are
assuming that the directions $x^\alpha$ are space-like; otherwise we
shall need $\eta^{\alpha\beta}$), and we can define a new vacuum state
$|\Omega_b\rangle$ such that $a_0^\alpha |\Omega_b\rangle=0$.
The relation between the  momentum
basis and the new oscillator basis is
given by (for each string)
\be \label{e27}
|\{p^\alpha\}\rangle = (2\pi/b)^{-k/4} \exp \Bigl[-{b\over 4} p^\alpha
p^\alpha
+ \sqrt b a^{\alpha\dagger}_0 p^\alpha - {1\over 2} a^{\alpha\dagger}_0
a^{\alpha\dagger}_0 \Bigr] |\Omega_b\rangle\, .
\ee
In the above equation $\{p^\alpha\}$ label momentum
eigenvalues.
Substituting eq.\refb{e27} into eq.\refb{e99}, and integrating over
$p_{(i)}^\alpha$, we can express the three
string vertex as
\ben \label{e288}
|V_3 \rangle &=&
\int d^{26-k}p_{(1)} d^{26-k}p_{(2)} d^{26-k}p_{(3)}
\delta^{(26-k)}(p_{(1)} + p_{(2)} +
p_{(3)}) \nonumber \\
&& \hskip-13pt\exp\Bigl(-{1\over 2}
\sum_{r, s\atop m, n \ge 1} 
\eta_{\bmu\bnu} a^{(r)\bmu\dagger}_m
V^{rs}_{mn} a_n^{(s)\bnu\dagger}
- \sum_{r,s\atop n\ge 1} 
\eta_{\bmu\bnu} p_{(r)}^{\bmu} V^{rs}_{0n}
a_n^{(s)\bnu\dagger}
-{1\over 2}\sum_r \eta_{\bmu\bnu} p_{(r)}^{\bmu}
V^{rr}_{00}
p_{(r)}^{\bnu} \Big) |0, p\rangle_{123} \nonumber \\
&& \otimes \, \bigg( {\sqrt 3 \over (2\pi b^3)^{1/4}}
(V^{rr}_{00}+{b\over 2})\bigg)^{-k}
\exp\Big(-{1\over 2}
\sum_{r,s\atop m, n \ge 0} 
 a^{(r)\alpha\dagger}_m V^{\prime rs}_{mn}
a_n^{(s)\alpha\dagger} \Big)
|\Omega_b\rangle_{123} \, . 
\een
In this expression the sums over $\bmu, \bnu$ run from 0 to $(25-k)$,
and the sum
over $\alpha$ runs from $(26-k)$ to 25. Note that in the last line the
sums over $m$, $n$ run over 0, 1, 2 $\ldots$.
The new $b$-dependent $V'$ coefficients are
given in terms of the $V$ coefficients by
\ben \label{e299}
V^{\prime rs}_{mn}(b) &=& V^{rs}_{mn} - {1\over V^{rr}_{00} +{b\over 2}}
\, \sum_{t=1}^3 V^{tr}_{0m} V^{ts}_{0n} \, , \qquad m,n\ge 1\, ,
\nonumber
\\
V^{\prime rs}_{0n}(b) &=& V^{\prime sr}_{n0} = {1\over V^{rr}_{00}
+{b\over
2}}\, \sqrt{b} \,  V^{rs}_{0n}\, , \qquad n \ge 1\, ,\nonumber \\
V^{\prime rs}_{00}(b) &=& {1\over 3} {b\over V^{rr}_{00} +{b\over 2}}\,
,
\qquad r\ne s\, ,
\nonumber \\
V^{\prime rr}_{00}(b) &=& 1 - {2\over 3}
{b\over V^{rr}_{00} +{b\over 2}}\, .
\een
In deriving the above relations we have used eq.\refb{evcon}.
These relations can be readily inverted to find
\ben \label{e299a}
V^{ rs}_{mn} &=& V^{\prime rs}_{mn}(b)  + {2\over 3}\, {1\over 1-
V^{\prime rr}_{00}(b) }
\, \sum_{t=1}^3 V^{\prime rt}_{m0}(b)  V^{\prime ts}_{0n}(b) \, , \qquad
m,n\ge 1\, , \nonumber \\
V^{rs}_{0n} &=&
{2\over 3} \,\,  {1\over 1-
V^{\prime rr}_{00}(b)}\,  \sqrt{b} \,  V^{\prime rs}_{0n}(b)\, ,
\qquad n \ge
1\, ,\nonumber \\ V^{rr}_{00} &=&{b\over 6} \,{1+ 3 V^{\prime
rr}_{00}(b)
\over  1- V^{\prime rr}_{00}(b)}\, .
\een

We shall now describe how our variables $V^{rs}_{mn}$ and $V^{\prime
rs}_{mn}$ are related to the variables introduced in
ref.\cite{gross-jevicki}. For this we begin by comparing the variables
in
the
oscillator representation.
Since
ref.\cite{gross-jevicki} uses the $\alpha'=1/2$
convention rather than the $\alpha'=1$ convention used here, every
factor of $p$ ($x$) in \cite{gross-jevicki} should be multiplied
(divided)
by $\sqrt{2\alpha'}$, and then $\alpha'$ should be set equal to one in
order to compare with our equations.
With this prescription
eqs.(2.5b) of \cite{gross-jevicki} giving $a_0 = {1\over 2} \hat p -
i\hat x$
becomes $a_0 =
{1\over \sqrt{2}}\, \hat p
- {i\over \sqrt{2}} \hat x$,
which
corresponds
to our \refb{e266} for $b=2$. Thus, we can directly compare our
variables with those of
\cite{gross-jevicki} for the case $b=2$.

Ref.\cite{gross-jevicki} introduced a matrix $U$ which appears, for
example,
in their eq.(2.47). We shall denote this matrix by $U^{gj}$. This matrix
appears in  the construction of the vertex in the
oscillator basis (\cite{gross-jevicki}, eqn.(2.52) and (2.53)).
This implies that the $V'$ coefficients for $b=2$ can be expressed
in terms of $U^{gj}$ using their results. In particular, defining
$V^{\prime
r s}$ to be the matrices
$V^{\prime r s}_{mn}$ with
$m,n$ now running from 0 to $\infty$, we have
(see \cite{gross-jevicki}, 
eqn.(2.53)):\footnote{As explained
at the end of appendix \ref{a1}, $U^{gj}$
should really be identified with $\bar U$ of
ref.\cite{gross-jevicki}.}
\be \label{e300}
V^{\prime\, r\, s}(2) = {1\over 3} ( C' + \omega^{s-r} U^{gj} +
\omega^{r-s} \wb U^{gj})\,,
\ee
where $\omega = \exp (2\pi i/3)$,  $C'_{mn}=(-1)^m\delta_{mn}$
with $m,n\ge 0$,
and the matrix
$U^{gj}$ satisfies the relations
(eq.(2.51) of \cite{gross-jevicki}):
\be
\label{ident}
U^{gj\,\dagger} = U^{gj} \,, \quad \wb U^{gj} \equiv (U^{gj})^* = C'
U^{gj}
C' \,, \quad  U^{gj} U^{gj} = 1 \,.
\ee

Eq.\refb{e300} 
gives us,
$V^{\prime\, rr}_{00}(2)
= {1\over 3}(1 + 2 U^{gj}_{00})$. With this result, the last
equation in \refb{e299a} can be used with $b=2$ to find
\be
V^{rr}_{00} = {1 + U^{gj}_{00}\over 1 - U^{gj}_{00}}\, .  
\ee
Similarly, the second equation in \refb{e299a} gives:
\be
V^{rs}_{0n}  = {1\over 1-
U^{gj}_{00}}\,  \sqrt{2} \,  V^{\prime rs}_{0n}(2)\, , \qquad \hbox{for}
\quad n\ge 1\, .
\ee
Making use of \refb{e300} and $\wb U^{gj}_{0n}
= (U^{gj}_{0n})^*$ 
we find that we can write, for $n\ge 1$:
\be \label{ex1}  
V^{r\, s}_{0n} = {1\over 3} (\omega^{s-r} W_n + \omega^{r-s}  W_n^*)
\, ,
\ee
where
\be
\label{ewn}
W_n = {\sqrt{2}
U^{gj}_{0n} \over 1 - U^{gj}_{00}}
\,.
\ee
The first equation in \refb{e299a} together with \refb{e300}
gives us 
\cite{0008252}
\be \label{e11}
V^{r\, s} = {1\over 3} ( C +\omega^{s-r} U + \omega^{r-s} \wb U)\, ,
\ee
where $V^{rs}$, $U$ and $C$ are regarded as matrices with indices
running
over $m,n \geq 1$,
$C_{mn} = (-1)^m \delta_{mn}$ and $U$ is given as
\be
 U_{mn} = U^{gj}_{mn} +
{U^{gj}_{m0}U^{gj}_{0n} \over 1 - U^{gj}_{00}}\,.
\ee
By virtue of this relation, and the identities in \refb{ident}
we have that the matrix $U$ satisfies
\be \label{e133}
\wb U \equiv U^*= C U C, \qquad U^2 = \wb U^2 = 1, \qquad U^\dagger =
U\,
,\quad \wb U^\dagger = \wb U\, .
\ee
It follows from  
\refb{ident} and \refb{ewn} that $W_n$
satisfies the relations:
\be \label{ex22} 
W_n^* = (-1)^n W_n, \qquad \sum_{n\ge 1} W_n U_{np} = W_p, \qquad
\sum_{m\ge 1} W_m^* W_m = 2
V^{rr}_{00}\, .
\ee

\medskip
Finally, using \refb{e299}, \refb{ex1}, \refb{e11}, the coefficients
$V'(b)$ for arbitrary
$b$ can be made into matrices with $m,n\geq 0$, and, just as for
the case $b=2$ in \refb{e300},  can be
written as
\be \label{e30}
V^{\prime\, r\, s}(b) = {1\over 3} ( C' +\omega^{s-r} U' +\omega^{r-s}
\wb U')\,,
\ee
where
\ben \label{ex33}
U'_{00} &=& 1 - {b\over V^{rr}_{00} + {b\over 2}} \, , \nonumber \\
U'_{0n} &=& (U'_{n0})^* = {\sqrt b \over V^{rr}_{00} + {b\over 2}} \,
W_n
\,
, \quad n\geq 1\, , \nonumber \\
U'_{mn} &=& U_{mn} - {W_m^* W_n \over V^{rr}_{00} + {b\over 2}} \,
\quad m,n \geq 1 .
\een
Using eqns.\refb{e133} and \refb{ex22} one 
can show that
$U'$,
$\wb U' \equiv U'^*$ viewed as a matrix with $m,n\geq 0$
satisfies the relations:
\be \label{e311}
\wb U' = C' U' C', \qquad U^{\prime 2} = \wb U^{\prime 2} = 1, \qquad
U^{\prime \dagger} = U'\, .
\ee

\sectiono{On the $b$-dependence of the lump solution} \label{a2} 

In this appendix we shall address the question as to whether the
apparent
$b$-dependence of the lump solution given in eq.\refb{e32} could be
a gauge
artifact. In order to avoid cluttering up the formul\ae\ we shall focus
on
the matter part associated with a single direction transverse to the
lump:
\be \label{eb1}
|\Psi'\rangle\equiv \NN' \exp(-{1\over 2}
\sum_{m,n\ge 0} S'_{mn}
a^{\dagger}_m
a^{\dagger}_n) |\Omega_b\rangle\, ,
\ee
where
\be \label{eb2}
\NN' = \bigg( {\sqrt 3 \over (2\pi b^3)^{1/4}}
(V^{rr}_{00}+{b\over 2})\bigg)\{\det(1 - X')^{1/2}\det(1+T')^{1/2} \}\,
.
\ee
Since all states
under discussion are in the matter sector associated with this single
direction, we shall refrain from adding the
subscript $m$ to
various states and the $*$ operation.
The $b$ dependence of the state given above comes from four sources:
\begin{enumerate}
\item $b$ dependence of $\NN'$ (including implicit $b$ dependence of
$X'$ and $T'$ through eqs.\refb{e299}, \refb{e33}, \refb{e34}).
\item $b$ dependence of $S'_{mn}$ through eqs.\refb{e299}, \refb{e33},
\refb{e34}.
\item $b$ dependence of $a_0^\dagger$ through eq.\refb{e266a}. Under an
infinitesimal change in $b$, eq.\refb{e266a} gives
\be \label{eb3}
\delta a_0^\dagger = {\delta b\over 2 b} \, a_0, \qquad
\delta a_0 = {\delta b\over 2 b} \, a_0^\dagger\, .
\ee

\item $b$ dependence of $|\Omega_b\rangle$ due to the change in the
definitions of $a_0$, $a_0^\dagger$. 
Requiring that $(a_0 + \delta a_0)$
annihilates $|\Omega_b\rangle + \delta|\Omega_b\rangle$ gives
\be \label{eb4}
\delta|\Omega_b\rangle = - {\delta b\over 4 b} (a_0^\dagger)^2
|\Omega_b\rangle\, .
\ee

\end{enumerate}

A straightforward calculation 
(involving expansion of the exponential to
first order in $\delta b$ using the Baker-Campbell-Hausdorff formula)
gives:
\be \label{eb5}
\delta |\Psi'\rangle = \Big(\delta \ln \NN' - {1\over 4}\, {\delta
b\over
b}
\, S'_{00}\Big) |\Psi'\rangle - {1\over 2} \Big\{ \delta S'_{mn} +
{\delta b\over 2 b} (- S'_{0m} S'_{0n} + \delta_{0m}\delta_{0n}) \Big\}
a_m^\dagger a_n^\dagger |\Psi'\rangle\, .
\ee
We would now like to ask if the expression for $\delta|\Psi'\rangle$
given
above
can be represented as a gauge transformation of the
kind given in eq.\refb{en4}  
\be \label{eb6}
\delta_{gauge} |\Psi'\rangle = |\Psi' *^m \Lambda\rangle - |\Lambda
*^m\Psi'\rangle \, ,  
\ee
for some state $|\Lambda\rangle$ in the matter sector. Eq.\refb{eb5}
suggests that we look for a
$|\Lambda\rangle$ of the form: \be \label{eb7}
|\Lambda\rangle = \Lambda_{mn} a_m^\dagger a_n^\dagger |\Psi'\rangle\, .
\ee
(Note that we could have included a term in $|\Lambda\rangle$
proportional
to $|\Psi'\rangle$, but this does not contribute to the gauge
transformation of $|\Psi'\rangle$.) Using eq.\refb{e8} and the
general formula \refb{egen}
together with the identity
\ben \label{eb8}
&& \langle 0 | \exp\Big(\lambda_i a_i -{1\over 2} P_{ij} a_i a_j\Big)
a_p a_q
\exp\Big(\mu_i a^\dagger_i -{1\over 2} Q_{ij} a^\dagger_i
a^\dagger_j\Big)
|0\rangle
\nonumber \\
 &=& {\p^2\over \p\lambda_p\p\lambda_q} \bigg(\langle 0 |
\exp\Big(\lambda_i a_i
-{1\over 2} P_{ij} a_i a_j\Big)
\exp\Big(\mu_i a^\dagger_i -{1\over 2} Q_{ij} a^\dagger_i
a^\dagger_j\Big)
|0\rangle \bigg)\, ,
\een
we can show that
\ben \label{eb9}
\delta_{gauge} |\Psi'\rangle &=& -Tr( \BB \VV'(1 - \Sigma'\VV')^{-1} )
\, |\Psi'\rangle \nonumber \\
&&\hskip-12pt +  \Big\{ \pmatrix{V^{\prime 31} , 
& \hskip-5ptV^{\prime 32} }
(1 -  
\Sigma'\VV')^{-1} \BB (1 - \VV'\Sigma')^{-1}\pmatrix{V^{\prime 13}\cr
V^{\prime 23}} \Big\}_{mn}   a_m^\dagger a_n^\dagger |\Psi'\rangle\, , 
\een
where
\be \label{eb10} 
\BB = \pmatrix{-C'\Lambda C' & 0\cr 0 & C'\Lambda C'}, 
\qquad \Sigma' = \pmatrix{S' &
0\cr 0 & S'},
\qquad \VV' = \pmatrix{V^{\prime 11}
& V^{\prime 12}
\cr V^{\prime 21} & V^{\prime 22}}\, .
\ee
Thus requiring that $\delta_{gauge} |\Psi'\rangle$ is equal to
$\delta
|\Psi'\rangle$ given in eq.\refb{eb5} gives,
\be \label{eb11}
Tr( \BB \VV'(1 - \Sigma'\VV')^{-1} ) = -\Big(\delta \ln \NN' - {1\over
4}\, {\delta b\over b} \, S'_{00}\Big) \, ,
\ee
and
\ben \label{eb12}
&& \Big\{ \pmatrix{V^{\prime 31}\, , &\hskip-6pt  V^{\prime 32} }
(1 -  
\Sigma'\VV')^{-1} \BB (1 - \VV'\Sigma')^{-1}\pmatrix{V^{\prime 13}\cr
V^{\prime 23}}\Big\}_{mn} \nonumber \\
&=&  - {1\over 2} \Big\{ \delta
S'_{mn} +
{\delta b\over 2 b} (- S'_{0m} S'_{0n} + \delta_{0m}\delta_{0n})
\Big\}\, . \nonumber \\
\een
These give a set of linear equations for $\Lambda_{mn}$. In order to
show
that the change in $b$ corresponds to a gauge transformation, we need to
show the existence of $\Lambda_{mn}$ satisfying these equations.
Although
we do not have a proof of this, the fact that $\delta |\Psi'\rangle$
given
in eq.\refb{eb5} satisfies
the various consistency relations ({\it
e.g.}
$|\Psi' * \delta \Psi'\rangle + |\delta \Psi'*\Psi'\rangle
=0$, which follows from the fact that $|\Psi'\rangle+\delta
|\Psi'\rangle$ satisfies eq.\refb{e5})
gives us hope that the solutions to these equations do exist.

\sectiono{Conservation laws for the sliver } \label{a3}  

In this section we derive conservation laws obeyed
by the sliver  state, 
following the methods of
\cite{0006240}.
The surface state $\WWB$ is defined by the conformal map
\be
\tilde z = \tan^{-1} (z)  \quad \to \quad {d\tilde z \over dz}
= {1\over 1+z^2},
\quad  S(\tilde z, z) = - {2\over (1+z^2)^2}  \,.
\ee
Here $\tilde z$ is the global coordinate on the
once punctured upper half plane 
Im $(\tilde z)>0$ 
with the punture at $\tilde z=0$, 
and $z$ is the local
coordinate
around the puncture. $S(\tilde z, z)$ denotes the standard Schwartzian
derivative.

To obtain Virasoro conservations laws, we consider a globally defined
vector field $\tilde v(\tilde z)$, holomorphic everywhere except
for a possible pole at the puncture $\tilde z=0$. The standard
contour deformation argument then gives \cite{0006240}
\be \label{conscon}
\WWB \oint  d z \, v ( z) \left( T(z) -\frac{c}{12} S(\tilde z, z)
\right)  =   0\, ,
\ee
where contour integration is around the origin,
and $v(z)=\tilde v(\tilde z) \left( \frac{d \tilde z}{dz}\right)^{-1}$.
Taking $\tilde v (\tilde z) = \tilde z^p$  (with
$p\leq 2$ so that there is no pole at $\tilde z = \infty$), we get
\be
v(z) = [\tan^{-1} (z) ]^p (1+z^2) \,.
\ee
We can write explicitly the first few conservation laws:
\ben
&& \WWB \Bigl( L_1 + {1\over 3}L_3 - {7 \over 45}L_5 + \cdots \Bigr)
=0\\
&& \WWB \Bigl( L_0 + {2\over 3}L_2 - {2 \over 15} L_4+{2 \over 35}L_6+
\cdots \Bigr)
=0\\
&& \WWB \Bigl( L_{-1} + L_1  \Bigr)
=0\\
&& \WWB \Bigl( L_{-2}+ {c\over 6}  - {29 \over 45}L_2
+{128 \over
945} L_4 - {848\over 14175} L_6 + \cdots \Bigr) =0\\
&& \WWB \Bigl( L_{-3} + {61 \over 189}L_3
-\frac{2176}{14175}
L_5+ \cdots
\Bigr) =0\\
&&
\WWB \Bigl( L_{-4} - {c\over 3} + {608\over 945}L_2 - {629 \over
4725}L_4 + \frac{1312}{22275} L_6 + \cdots
\Bigr) =0 \,.
\een
Here $L_m$ could stand
for either matter or ghost Virasoro operators
(or
even the Virasoro operators associated with a subsector of the matter
conformal field theory) and $c$ is the corresponding central charge.
The perfect conservation of $K_1 = L_1 + L_{-1}$
holds more generally for
any ``wedge state'' $\langle n |$,
defined by the conformal map (see \refb{emapn})  
\be  
\tilde z = \frac{n}{2} \, \tan \left(\frac{2}{n} \tan^{-1} z \right) \,.
\ee
Indeed, with $\tilde v(\tilde z) =\tilde z^2 +1$ we get $v(z) =
(z^2 +1)$
and deduce that for all $n$,
\be
\langle n | K_1 =0 \,.
\ee
The operator $K_1$ is a derivation of the $*$ product, and
the observation that it annihilates all wedge states fits nicely
with the fact that the wedge states form a subalgebra under
$*$ multiplication.

\bigskip

Conservation laws for the antighost $b$ are identical to Virasoro
conservations with vanishing central charge,  since $b$ is a
true conformal primary of dimension two.

\bigskip

Conservations laws involving $c_n$'s
can be derived analogously.
Here we need to consider holomorphic quadratic differentials
$\tilde\varphi
(\tilde z)$.  
We can take $\tilde\varphi = 1/\tilde z^p$, with $p\geq 4$
for the quadratic differential to be regular
at $\tilde z \to \infty$. Back in $z$ coordinates one has
\be
\varphi (z) =\left( \frac{d \tilde z}{dz}\right)^2  
\tilde \varphi(\tilde z) =
{1\over [\tan^{-1}
(z) ]^p
(1+z^2)^2} \,.
\ee
Contour deformation argument now gives 
\be
\WWB \oint  d z \,\varphi ( z) c(z)  =   0 
\,.
\ee
The first few 
conservations read
\ben
&& \WWB \Bigl( c_{-2} - {2c_0\over 3} + {29 c_2\over 45} - {608c_4\over
945} +\cdots
\Bigr) =0\\
&& \WWB \Bigl( c_{-3} - {c_{-1}\over 3} + { c_1\over 3} - {61 c_3\over
189} +\cdots
\Bigr) =0\\
&& \WWB \Bigl(c_{-4} + {2c_{0}\over 15} - {128 c_2\over 945} + {629
c_4\over 4725}
+\cdots
\Bigr) =0\\
&& \WWB \Bigl( c_{-5} + {7c_{-1}\over 45} - {7 c_1\over 45} + {2176
c_3\over
14175} +\cdots
\Bigr) =0\\
&& \WWB \Bigl( c_{-6} - {2c_{0}\over 35} +
{848 c_2\over 14175} - {1312 c_4\over
22275} +\cdots
\Bigr) =0 \,.
\een

\end{document}